\documentclass[10pt]{article}
\usepackage{epsfig,graphicx,cite}
\usepackage{times,ch2005}

\def\beq{\begin{equation}}
\def\eeq#1{\label{#1}\end{equation}}
\def\eeqn{\end{equation}}

\def\beqa{\begin{eqnarray}}
\def\eeqa#1{\label{#1}\end{eqnarray}}
\def\eeqan{\end{eqnarray}}

\def\Dslash{\not{\hbox{\kern-4pt $D$}}}
\def\dslash{\not{\hbox{\kern-2pt $\del$}}}

\newcommand{\tev}{\ensuremath{\mathrm{\,Te\kern -0.1em V}}\xspace}
\newcommand{\gev}{\ensuremath{\mathrm{\,Ge\kern -0.1em V}}\xspace}
\newcommand{\mev}{\ensuremath{\mathrm{\,Me\kern -0.1em V}}\xspace}
\newcommand{\kev}{\ensuremath{\mathrm{\,ke\kern -0.1em V}}\xspace}
\newcommand{\ev}{\ensuremath{\mathrm{\,e\kern -0.1em V}}\xspace}
\newcommand{\gevc}{\ensuremath{{\mathrm{\,Ge\kern -0.1em V\!/}c}}\xspace}
\newcommand{\mevc}{\ensuremath{{\mathrm{\,Me\kern -0.1em V\!/}c}}\xspace}
\newcommand{\gevcc}{\ensuremath{{\mathrm{\,Ge\kern -0.1em V\!/}c^2}}\xspace}
\newcommand{\mevcc}{\ensuremath{{\mathrm{\,Me\kern -0.1em V\!/}c^2}}\xspace}

\def\mus  {\ensuremath{\rm \,\mus}\xspace}

\def\mus        {\ensuremath{\,\mu{\rm s}}\xspace}    

\begin{document}


\Title{Next generation of IACT arrays: \\
\em \large  scientific objectives versus  energy domains}
\bigskip


%
\label{AharonianStart}

%
\author{Felix Aharonian\index{Aharonian, F.} }

%
\address{Max-Planck-Institut f\"ur Kernphysik\\
Saupfercheckweg 1, 69117 Heidelberg, Germany \\
}

\makeauthor\abstracts{Several key motivations and perspectives
of ground based gamma-ray astronomy are discussed
in the context of the specifics of detection techniques and 
scientific topics/objectives  relevant to four major 
energy domains -- very-low or  \textit{multi-GeV}  ($E \leq$ 30~GeV), 
low or \textit{sub-TeV} (30~GeV - 300~GeV), 
high or \textit{TeV} (300~GeV - 30~TeV), and very-high or \textit{sub-PeV}
($E \geq$ 30~TeV)  intervals --  to be covered 
by the next generation of IACT arrays.}  

\section{Introduction}

The recent success of ground-based gamma-ray astronomy, 
in particular the exciting discoveries of many new galactic and 
extragalactic  TeV $\gamma$-ray emitters by \textit{HESS} 
(see e.g.  Ref.\cite{Hofmann-ch2005}),  elevated the status of the field
from an "astronomy with several sources" to the level
of truly observational discipline. In addition to  the important
 astrophysical and cosmological implications, these results will 
have a considerable  impact on  plans of the gamma-ray community 
towards the next generation of Imaging Atmospheric Cherenkov 
Telescope (IACT) arrays. One of the principal issues in this regard 
is the choice of the  energy domain. 
The  imaging of air showers with atmospheric  Cherenkov telescopes\cite{Jelley_Porter,Weekes_Turver,Stepanian,Hillas1,Hillas2,Weekes2,
Fegan,Hoffman}, 
especially in the stereoscopic mode is a powerful detection 
technique\cite{Aharonian93,AhAk:97,Aharonian97} 
which (potentially) allows  coverage of a  broad energy range
extending from  $\leq $ 10 GeV 
to $\geq $ 100 TeV $\gamma$-rays \footnote{Note that the first attempt 
of stereoscopic observations of air showers in 
the early 1970s using the so-called "double-beam"  
method\cite{Grindlay} actually  contained (indirectly)  also 
some  elements of the modern  imaging Cherenkov technique.}.

If one limits the energy region to a relatively 
modest threshold around 100 GeV, the performance
of the telescope arrays and their implementation
can be predicted with confidence. Therefore it was quite 
natural that in the mid 1990s  the stereoscopic IACT arrays  
consisting of 10m-diameter class reflectors were recognized, 
among a variety competing designs,
as the most effective approach that could  facilitate a 
qualitative improvement in performance at an affordable cost 
and with a guaranteed fast scientific return  \cite{AhAk:97}.  
This choice was soon accomplished in the form 
of the \textit{HESS} 4-telescope (Phase-1) array.
The results obtained during the first two years of operation 
of this array fully support  the most optimistic 
predictions concerning both the instrumental performance and the 
astrophysical implications, in particular,  the high quality 
morphological and spectrometric studies of extended 
galactic regions in TeV $\gamma$-rays.  
It is expected that together with 
two other similar projects - \textit{CANGAROO-III} \cite{Mori-ch2005}
and  \textit{VERITAS} \cite{Weekes-ch2005}, 
as well as the \textit{MAGIC} system consisting  of two 
17m-diameter telescopes \cite{Moriotti-ch2005},
the \textit{HESS} (Phase-1) array  will dominate the field for the 
next several years (see e.g. Ref.\cite{Weekes3}). 
At the same time, the  great success of \textit{HESS}  
supplies a strong rationale for the next generation 
IACT arrays.       

\section{Planning future IACT Arrays}
The ultimate goal in planning of the next  
generation IACT arrays should be a dramatic  
(down to the level of $10^{-14} \ \rm erg/cm^{2} s$) 
improvement of the flux sensitivity in the 
classical/standard (0.1-10 TeV) energy regime, 
and an aggressive expansion of the energy 
domain of IACT arrays in two directions -
down to (multi)GeV energies and  up to (sub)PeV energies.   
In this regard, I believe that  the  design
studies of future IACT arrays will proceed in four independent, 
although tightly correlated  and complementary  
directions with an ambitious aim to cover 
(more or less) homogeneously a very broad  energy range 
extending from $\leq 0.03$ TeV (30~GeV) to $\geq 300$ TeV (0.3~PeV). 
Below I discuss some basic requirements in the following  
four energy regimes in the context 
of detection specifics and principal scientific issues 
to be addressed by future IACT  arrays:

{
~~~~$\bullet$ very-low or  \textbf{multi-GeV} : ~$\leq$ 30~GeV 

~~~~$\bullet$ low or          \textbf{sub-TeV} : ~~~~~~~~~~~~~30~GeV - 300~GeV

~~~~$\bullet$  high or      \textbf{TeV } : ~~~~~~~~~~~~~~~~~300~GeV - 30~TeV
 
~~~~$\bullet$  very-high  or  \textbf{sub-PeV}  : ~~~$\geq$ 30~TeV 
}

\subsection{TeV Regime: {\sf  $10^{-14} \  \rm erg/cm^{2} s$} \sl 
sensitivity IACT Arrays} 
This is the most natural/intrinsic for the IACT technique
energy regime, where the combination of three basic factors -- (i) the 
high efficiency of detection/identification of electromagnetic 
showers, (ii) good accuracy of reconstruction of the direction 
and energy of primary $\gamma$-rays, and (iii) the large $\gamma$-ray photon 
statistics --  allows the best energy flux sensitivity at   
the level of $10^{-14} \ \rm erg/cm^{2} s$
($\approx  0.3$ milliCrab at 1 TeV),  and angular resolution 
$\delta \theta  \approx$ 1-2 arcminutes. This can be achieved by
stereoscopic arrays consisting of  multi (up to 100 or so) 
10m-diameter class (HESS-type) telescopes.

The flux sensitivity $10^{-14} \rm \ erg/cm^2 s$ 
at TeV energies would be a great and impressive 
achievement even 
in the standards of the most advanced 
branches of observational astronomy. This should allow us to probe 
the $\gamma$-ray luminosities of potential TeV emitters at the 
levels of $10^{32} (d/10 \rm kpc)^2 \ \rm erg/s$ for galactic sources and 
$10^{40} (d/100 \rm Mpc)^2 \ \rm erg/s$ for extragalactic objects. 
Although for moderately extended sources, e.g. of  angular size 
$\Psi \sim 1^\circ$, the minimum detectable energy flux will be by
a factor of $\Psi / \delta \theta \sim$10-30 
higher, yet it would  be better than the 
energy flux sensitivities of the best current X-ray satellites, \textit{Chandra}, 
\textit{XMM}-Newton  and \textit{Suzaku},  in  the keV band, i.e. should allow the 
deepest probes of nonthermal high energy phenomena in 
extended sources, in particular in shell type Supernova Remnants (SNRs),
Giant Molecular Clouds (GMCs), 
Pulsar Driven Nebulae (Plerions), Clusters of Galaxies, hypothetical 
Giant Pair Halos around AGN, \textit{etc.}

Therefore,  one of the prime objectives and urgent issues  of 
ground-based gamma-ray astronomy in 
the foreseeable future should be the 
design and construction of a  
"$10^{-14} \ \rm erg/cm^{2} s$ \textit{sensitivity IACT Array}".  
Since this will be 
essentially a TeV instrument with a limited capability for study of 
extragalactic objects located beyond  $z=$0.2-0.3, 
but with a great potential for morphological and spectrometric measurements and deep surveys, 
this array  should be dedicated, first of all, to 
studies of galactic sources  and  the diffuse emission of the Galactic 
Disk.

The $10^{-14} \ \rm erg/cm^{2} s$ flux sensitivity and 
$\geq 5^\circ$ FoV should provide very deep surveys of 
the Galactic Disk, as well as of some selected regions 
above the Galactic Plane. One may predict, based on the 
extrapolation of \textit{HESS} results, that 
such an instrument should discover and resolve 
hundreds, or perhaps even thousands of galactic 
TeV  sources. In particular, this  array should allow statistically significant 
detection of the weakest \textit{HESS} sources for exposure times 
less than 1 hour, and, more importantly, 
detailed studies of spectral and spatial structures 
of relatively strong  ($\geq 1$ milliCrab)  \textit{HESS } sources. 
This could provide key insight into the origin of Galactic 
Cosmic Rays (GCRs), in particular  decisive tests for 
the hypothesis that  shell type SNRs are responsible for 
the \textit{bulk} of 
observed cosmic rays  up to $10^{15} \ \rm eV$. Although several
shell type SNRs already have been reported as TeV emitters,
and in all cases the TeV emission can be explained 
quite naturally by hadronic ($pp$) interactions, 
 the limited information about  both the spectral and spatial 
distributions of detected signals does not allow definite conclusions
concerning the nature  of TeV emission,  especially  
because the latter could be 
substantially ``contaminated'' by $\gamma$-rays  
of leptonic (inverse Compton) origin. Moreover
the evidence  of the hadronic origin of TeV emission
from a few selected SNRs does not yet imply that the bulk of 
GCRs can be explained by shell type SNRs. 
The discovery \textit{by HESS} of TeV $\gamma$-rays
from several plerions and two binary systems (of different origin)
indicates that other galactic sources, 
in particular pulsars/pulsar-winds  and microquasars, may 
contribute  comparably to  the flux of locally observed  cosmic rays. 

While young particle accelerators can be 
identified directly, i.e.  through their characteristic  $\gamma$-ray
emission, in the case of old sources in which the particle 
acceleration has ceased and the most energetic (TeV and PeV) particles 
have already left the source, the TeV $\gamma$-ray emission should be 
significantly suppressed. On the other hand, 
one may expect  detectable TeV $\gamma$-ray 
emission from interactions of \textit{run-away protons} with the nearby 
dense environments. Giant Molecular Clouds (GMCs)   with 
diffuse masses $10^4$ to  $10^6 \ \rm M_\odot$, seem 
to be ideal objects to serve as effective  targets.     
These  objects are intimately connected with star formation regions  
that are strongly believed to be the most probable locations 
(with or without SNRs) of cosmic ray production in our
Galaxy. The search for TeV photons from GMCs is important to ascertain the 
possible existence of nearby high energy proton accelerators.    
Remarkably, $10^{-14} \ \rm erg/cm^2 s$ sensitivity 
should allow  detection of TeV gamma-ray emission not only from 
GMCs located close to young particle  accelerators, but also from "passive 
clouds" located in ordinary sites of interstellar medium, 
far from  active accelerators. Such clouds can serve as unique 
"barometers" for measuring the pressure (energy density)
of protons of the sea of GCRs in different parts of the Galactic Disk.
Since the hadronic component of 
the  diffuse $\gamma$-ray background
is essentially contributed by individual clouds, 
the potential of the $10^{-14} \ \rm erg/cm^2 s$ 
sensitivity arrays  to resolve the small-scale ($\sim 0.1^\circ$)
features in the form of individual $\gamma$-ray emitting clouds,
and thus to study the variations of GCRs on $\sim 10$ to 100 pc scales,
would be crucial for understanding of many aspects 
of the origin and propagation of galactic cosmic rays. 

This array will be a very effective tool also for spectrometric 
and temporal studies  of highly  variable phenomena in compact 
galactic objects like microquasars, as well as   
Sgr A* -  presumably a  Supermassive  Black Hole in the 
center of our Galaxy. To a large extent  this concerns also   
relatively nearby ($z \leq 0.1$) extragalactic sources, 
in particular BL Lac objects.  
In this regard one may predict the discovery of numerous  
BL Lacs, although with significantly absorbed TeV spectra.

Another aspect of extragalactic studies, with a  promise 
for exciting findings,  is connected to  
the search for several components of TeV radiation 
from different parts (compact cores, kpc-scale jets, 
radio lobes) of two nearby  radiogalaxies, M87 and Cen A,
as well as from luminous starburst galaxies like Arp 220. 
Finally, this array should  
detect TeV $\gamma$-ray emission expected from nearby 
clusters of Galaxies, like the Virgo and Coma clusters, unless our current 
understanding of the nonthermal 
energy budgets and acceleration processes in 
these unique cosmological reservoirs of cosmic rays is completely wrong.    
It is clear, however, that  because of the severe intergalactic absorption 
of  TeV $\gamma$-rays,  many extragalactic 
and cosmological topics of gamma-ray astronomy can be addressed
adequately  by lower energy-threshold instruments.  

\subsection{Sub-TeV Regime: \sl  Very Large Aperture IACT Arrays} 
The energy threshold $\varepsilon_{\rm th}$ of IACTs 
is generally defined as a characteristic energy  at which
the $\gamma$-ray detection rate for a primary  
power-law spectrum  with photon index  $\Gamma=$2-3 
achieves its maximum. It is well known, from Monte Carlo 
simulations and from the operation  of  
previous generation  IACTs, 
that in  practice the best performance, in particular 
the minimum detectable energy flux,   is 
achieved at energies exceeding several times  $\varepsilon_{\rm th}$. 
In this regard, for optimization of  $\gamma$-ray detection
around 100 GeV, one should  reduce the energy threshold 
of telescopes  to $\varepsilon_{\rm th} \leq 30 \ \rm GeV$. 
This can be done   
by using very large, $\geq 20$~m-diameter class reflectors and/or 
very high ($\geq 40 \%$)   quantum efficiency fast (ns) optical receivers.    
On the other hand,  reduction of the detection threshold to such 
low energies is an important  scientific issue in its own right; 
the intermediate interval between 30 and 300 GeV 
is a crucial energy regime for certain class of galactic 
and extragalactic $\gamma$-ray source populations.   

Within the next two years 
a stereoscopic system consisting of two MAGIC 
17m-diameter telescopes \cite{Teshima-ch2005}, 
will start to take meaningful probes from this 
important energy band. In  a  more distant future, 
this energy region  can  be comprehensively explored by
the \textit{HESS} and \textit{MAGIC} 
collaborations using 30m-diameter class 
telescope arrays; presently two prototypes of such extremely 
big telescopes are in the construction \cite{Punch-ch2005} or  
design-study \cite{Teshima-ch2005}  stages.  
 
In  order to achieve the lowest possible energy threshold,
the integration time of the Cherenkov light 
should be reduced to $\leq 10$ ns. Coupled with the 
requirement of high quality imaging, this  
constrains the choice of possible configurations 
of 30m-diameter class optical reflectors,
and correspondingly limits the FoV to $\sim 3^\circ$. 
While for point-like sources this is not a 
big disadvantage, it limits significantly the capability 
of these telescopes for  studies of  
extended sources. Also, it limits the $\gamma$-ray detection area 
at high energies. Therefore,  the  optimization of the  
performance for point-like sources in the 
energy  range  below 1 TeV should be  a key    
issue  in design studies of  future 30m-diameter telescope 
arrays.
In this sense the 30m-diameter class IACT arrays 
can be considered as essentially sub-TeV instruments. 

There are several  reasons to believe that the energy regime  
below 100 GeV should be prolific in number of $\gamma$-ray sources,
especially when the flux sensitivities  of IACT arrays 
in this energy band achieve the level close to  
$10^{-13} \ \rm erg/cm^2 s$. 
From a purely  phenomenological perspective,
TeV (\textit{HESS }) and GeV (\textit{EGRET}) 
sources are expected to show up in the immediate neighbor 
region between 10 and  100 GeV, even in the case of significant 
hardening  of the spectra of TeV sources at lower energies,
or  strong steepening of GeV sources at higher energies.
In particular,  because of energy-dependent escape of protons and electrons,
as well as  severe radiative  losses of  ultrarelativistic electrons,  
one may expect much steeper spectra 
of particles  inside old objects compared to young, typically  
1000 yr old accelerators.  If so, the major fraction of 
galactic accelerators should have steep proton spectra, 
and therefore  higher chances to be detected in 
$\gamma$-rays  at  energies below $\leq 100$ GeV. 

The $\gamma$-ray sources with steep energy spectra 
should dominate, although for different reasons, 
e.g. due to the intergalactic and internal
photon-photon  absorption, also in   
extragalactic source populations. Therefore, 
while several blazars with very steep energy spectra
have been detected by \textit{HESS } after   
significant efforts (long exposure times), they should  be detected  
much easier at energies below 100 GeV. 
Thus,   in this energy band we  may expect  a 
dramatic increase of the number of detectable 
blazars  up to redshifts $z \sim 1$. 

Generally, the 30-300 GeV and 300 GeV-30 TeV energy bands
share many common phenomena and sources. In this regard,
the studies below 100 GeV are 
complementary, in some cases even crucial, for understanding 
the origin of $\gamma$-ray emission. For example, 
the spectral measurements in this energy interval may 
provide decisive information whether the TeV emission
from shell type SNRs and plerions has inverse Compton 
(very hard energy spectra at low energies, 
with gradual steepening above 1 TeV) or $\pi^0$-decay 
(typically  single power-laws throughout the 10 GeV to 
$\geq 10$ TeV interval) origin. 
Another example. In compact leptonic sources 
(e.g. in binary systems) one should expect
transition of the process of   $\gamma$-ray production 
from the Thompson to the Klein-Nishina regime or 
(in the case of binaries with  very luminous optical companion stars) 
from optically thin (at GeV energies) to optically thick 
(at TeV energies) regimes. The corresponding characteristic 
spectral features in the 10 GeV  to  1 TeV interval 
can serve as distinct signatures of $\gamma$-ray production regions.

\subsection{Sub-PeV Regime: \sl  10km$^2$ IACT Arrays} 

The current trend to reduce  the energy threshold of 
ground-based  detection technique concerns both 
the atmospheric Cherenkov telescopes and the air-shower
(particle detector)  arrays.  As a result, presently there is  only little  
instrumental  activity   in the  energy domain above  30~TeV.   
The general tendency of decreasing   $\gamma$-ray fluxes  
with energy becomes especially dramatic above 10 TeV. 
The reasons could be  different, e.g. external and internal 
absorption of $\gamma$-rays, limited efficiency
of particle acceleration processes, escape of highest energy 
particles from the production region, \textit{etc.}  
This limits the capability of traditional air-shower 
arrays, mainly   because of the  limited proton/gamma separation power,  
limited angular resolution and limited detection areas.
Any meaningful study of cosmic $\gamma$-rays beyond 
30 TeV requires detection areas exceeding  $1 \ \rm km^2$. 
At large zenith angles, $60^\circ$ or so, the collection area
of the atmospheric Cherenkov detectors increases rapidly. 
Thus the use of IACT systems at  large zenith angles can 
improve the $\gamma$-ray statistics 
in the multi-TeV or sub-PeV region. For  many astrophysical objects, 
on the other hand, the observation time in this mode is 
quite short  (because the sources set rapidly 
below  the horizon). Besides, even small variations of 
the atmospheric transparency add non-negligible
uncertainties  in the derivation of  
shower parameters obtained at large zenith angles.     

An  effective and straightforward approach    
would be  the use of  IACT arrays optimized 
for detection of $\gamma$-rays in the 30 to 300 TeV region.  
Such an array  can  consists  of rather modest,
approximately  20 to 50  m$^2$  area reflectors
separated from each other,   
depending on the scientific objectives and the configuration of 
the imagers,  between 300 to 500~m.
The  requirement  to  the  pixel size of  imagers is also 
quite modest,  between $0.25^\circ$ to $0.5^\circ$, however 
they should have large FoV  
$\ge  5^\circ$  in order to detect 
showers from distances $\geq 300$~m. 
It is expected  that 
an array consisting of several tens of such telescopes can provide 
an extraordinary large detection  area of about $10 \ \rm km^2$,
reasonable  efficiency for   suppression of 
hadronic showers,  very good  angular resolution of few arcmin and 
quite good, better than $0.15 \%$, 
energy resolution \cite{Plya:2000}.

Although the prime motivation of such an array is the 
energy region above 30 TeV, it can serve as an extremely 
powerful tool for detailed spectrometric and morphological
studies of lower (3 to 10 TeV) energy $\gamma$-rays as well.
In particular, all \textit{HESS} sources with hard 
energy spectra extending beyond several TeV, 
would be perfect targets for studies
with unprecedented TeV photon statistics.
  
But, of course, the highest priority scientific topics 
for such an instrument are  linked to the search and
studies of  "Cosmic PeVatrons"  and the surrounding  regions.
This concerns, first of all, galactic sources, in particular 
the shell type SNRs, Molecular Clouds, Pulsar Wind Nebulae
and Microquasars.  For example, the detection  
of $\geq 30$ TeV  $\gamma$-rays   from shell type SNRs would be 
direct proof  that the shocks in SNRs accelerate 
protons up to energies $10^3$ TeV. While one may hope to have
only handful SNRs  detected at energies 
above 30 TeV, namely the youngest ones
of age less than $\approx 1000$ year, the chances to detect 
$\gamma$-rays well beyond 10 TeV are significantly higher for 
molecular clouds located nearby the Galactic PeVatrons. 
Because of  the (suspected) slow diffusion of charged particles in 
turbulent environments in star formation regions,  
where the  cosmic accelerators appear most frequently, we 
should expect "delayed" emission from interactions of 
cosmic rays, which left their production sites (accelerators)  
some time  ago,  with 
nearby dense molecular clouds. The energy 
spectra of secondary $\gamma$-rays strongly depend            
on the diffusion coefficient of charged particles, the
age of the accelerator, and the distance between the 
accelerator and the cloud. For some combinations of 
relevant parameters (young source and/or slow diffusion and/or
distant clouds), one may expect extremely hard 
$\gamma$-ray spectra  with a maximum  (in the $\nu F_\nu$ plot) 
beyond 10 TeV.    

The synchrotron nebulae initiated by termination of 
relativistic pulsar winds, where  particles can be 
accelerated to much higher energies than in shell type SNRs, 
are expected as another prolific class of multi-TeV and sub-PeV
emitters.  

Because of severe intergalactic absorption
one may expect $\geq$ 30 TeV $\gamma$-rays only  
from a handful  extragalactic sources like 
radiogalaxies Cen A and  M87, as well as 
from some nearby starburst and normal  
galaxies.  Although the sub-PeV IACT arrays should be  
dedicated for galactic studies,
the detection of $\geq 30$ TeV $\gamma$-rays from 
nearby extragalactic objects would have, among other 
astrophysical implications,  a great 
cosmological importance 
for probing the far-infrared cosmic background. 
The distances to these objects perfectly match the 
$\gamma$-ray astronomical method  of deriving 
information about the  cosmic infrared background radiation 
at $\geq 30  \mu$ wavelengths.

\subsection{Multi-GeV
 Regime: \sl Gamma-Ray Timing Explorers} 
One of the greatest expectations of gamma-ray astronomy 
is connected with the launch of \textit{GLAST}. This  
instrument, with a nice  performance 
between 30 MeV to 10 GeV,  allows  also  
an extension of study  of the $\gamma$-ray sky
to 100 GeV.  Thus the gap between space-based and ground-based 
instruments will finally disappear.  Generally this 
is considered as one of the major  achievements  of 
observational gamma-ray astronomy. Although certain 
technical (cross-calibration of instruments) and 
scientific (broad band $\gamma$-ray coverage) 
aspects of this issue are quite important, 
the astrophysical significance of the expected 
overlap of detection domains of GLAST and the current  
ground-based instruments seems somewhat 
overemphasized in the literature. 
Although at GeV energies \textit{GLAST} will 
improve the \textit{EGRET} 
sensitivity by almost two 
orders of magnitude, the capability of \textit{GLAST} and, 
in fact, of  any post-GLAST space project  at energies 
beyond 10 GeV will be quite limited, first of all 
because of the limited detection area, 
unless the Moon would be used in (far) 
future as a possible platform for installation 
of very large, $\geq 100 \ \rm m^2$ area 
pair-conversion detectors.  It is clear that  
the space-based resources of GeV gamma-ray astronomy 
have achieved a point where any further progress 
would appear  extremely  difficult and  very expensive. 
For the next decades to come there is no space-based 
mission planned  for the exploration of the gamma-ray sky. 
 
The impressive sensitivity 
of \textit{GLAST} at 1 GeV -
few times $10^{-13} \ \rm erg/cm^2 s$ -      
can be achieved after  one year all-sky survey.  
While for the persistent $\gamma$-ray  sources
this is an adequate estimate of the performance
(taking into account  that 
a huge number of  sources  will be simultaneously monitored 
by  the large, almost $\sim 2 \pi$ steradian  homogeneous 
FoV), the small, $\approx 1 \ \rm m^2$   
detection area limits the potential of this instrument
at  GeV energies 
for detailed studies of the temporal and spectral 
characteristics of highly variable sources like blazars
or solitary events like  gamma-ray bursts (GRBs).  
In this regard, the need for a powerful 
multi-GeV instrument to study transient 
phenomena with adequate high energy $\gamma$-ray
photon statistics,  has motivated 
the idea/concept to extend  of the 
domain of the imaging atmospheric Cherenkov technique,
with its huge collection area $\geq 10^4 \ \rm  m^2$,  
down to energies of about 5 GeV. 
In practice, this can be achieved with a 
stereoscopic telescope system consisting of several  
$\geq 20$m diameter dishes  located at 
high elevations  of the order of 
5 km  above sea level. That is why the concept was called  
5@5 \cite{Aharonian:2001}.  
The successful  realization  of  such an instrument 
largely depends on the availability of 
sites with a 
dry and transparent  atmosphere at an altitude as high as 5 km. Nature does
provide us with such an extraordinary site - the  
Atacama desert in Northern Chile which  
has been chosen for the installation of one of the most powerful future astronomical 
instruments  -- the Atacama Large Millimeter Array (ALMA). 
Several good sites for installation of high-altitude 
multi-GeV telescope arrays exist also in the Northern 
Hemisphere, e.g. in India 
(Hanle, 4.2km asl) \cite{Koul-ch2005}. 
  
Another approach to achieve 
a sub-10 GeV  energy threshold, but at more comfortable  
altitudes,  is linked to very large aperture, 
30m diameter telescopes equipped with novel high quantum 
efficiency receivers. Clearly, the 
combination of 3 key elements --  high altitude, large 
optical reflectors, and  high quantum efficiency focal plane 
receivers -- would be an ideal combination for 
construction of $\gamma$-ray
detectors with an energy threshold as low as  several GeV.

Dramatic reduction of the energy threshold of detectors 
is a  key issue for a number of astrophysical and cosmological
problems, e.g. for study of $\gamma$-radiation from pulsars 
and cosmologically distant objects like quasars and GRBs.  
The  spectra of typical representatives of 
all three source populations contain, for different reasons, 
sharp energy cutoffs  around or below 10 GeV. That is why 
the \textit{uncompromised}  reduction of  the energy 
threshold of detectors down to several GeV is so critical, 
and would justify the choice of such uncomfortable altitudes 
for operation of large IACT arrays. 

The concept of 5@5 is not only motivated by 
the possibility of 
coverage of the yet unexplored region of multi-GeV $\gamma$-rays.
In fact, 5@5 combines two advantages of the current ground-based 
and satellite-borne $\gamma$-ray domains - large photon fluxes 
and enormous detection areas. 
This  would make the  5@5 a unique \textit{Gamma-ray Timing Explorer}
(with a sensitivity to detect the hard-spectra EGRET sources  
in exposure times  of  seconds to minutes)
for  the study of transient non-thermal $\gamma$-ray  phenomena like 
rapid variability of Blazars, synchrotron flares in Microquasars,
the high energy (GeV) counterparts of Gamma Ray Bursts, {\em etc.}.

If the spectra of GRBs extend to high energies, 
which is the case at least for some of GRBs,
then the sensitivity of  5@5 should allow 
detailed studies of spectral and temporal features
of GRBs in this extremely important energy
band.  The detection of $\geq 5$ GeV episodic events with 
 typical GRB fluxes 
$\geq 10^{-8} \, \rm erg/cm^2 s$ (at keV-MeV energies)  
would require $\leq 1 \, \rm s$ observation time. 
Thus it would be possible to monitor
the spectral evolution of  GRBs  on very short, sub-second timescales.  
 Remarkably, 
even for fluxes as low as $10^{-10} \, \rm erg/cm^2 s$, 
the required detection  time does not exceed 100 sec. Some 
of the GRB models predict multi GeV emission
during the  afterglow phase of 
evolution. If so,   5@5   could serve as a unique tool
for studies of  the properties of  GRBs  at  late stages of their 
evolution, and thus 
provide a key information  about these most mysterious objects 
in the Universe.

The capability of 5@5  is not limited by variable 
source studies. This instrument, in fact, will have significantly 
broader goals  related to detailed spectrometry in the 
multi-GeV energy band of  $\gamma$-ray sources 
like SNRs, pulsars, unshocked pulsar winds, 
perions, large (kpc)  scale extragalactic jets, clusters of galaxies, \textit{etc}. 
For example, the direct searches 
for pulsed GeV radiation  from a fraction of 
unidentified EGRET sources  (suspected to be pulsars) 
without invoking information from the longer 
(radio, optical, X-ray)  wavelengths, seems to be an 
important issue, especially  because in many pulsars 
the periodic signals at low  frequencies  could be suppressed. 

Finally, the reasonably good energy resolution in the 
energy interval between 10 and 100 GeV, coupled with  
adequate  gamma-ray photon statistics, is crucial for effective  
cosmological studies  through (1) probing the 
cosmological evolution of the cosmic 
background radiation at optical and UV wavelengths, 
(2) detecting $\gamma$-rays from large scale  
structures (Galaxy Clusters) , and 
(3) searching for characteristic emission from the 
non-baryonic Dark Matter Halos.

\section{Discussion}
Different classes of IACT arrays  described above  
are  characterized by  certain energy  intervals in which the  
best energy flux sensitivity  is  achieved.
These "best performance energy intervals" can be grouped 
in the following segments :  
$\leq 30 \ \rm GeV$ (\textit{multi-GeV} band),  
30 GeV - 300 GeV (\textit{sub-TeV} band),
0.3 TeV to 30 TeV (\textit{TeV} band), 
and  $\geq 30 \ \rm TeV$ (\textit{multi-TeV } or \textit{sub-PeV} band),
respectively.  It should be noted, however, that all four 
versions of IACT arrays allow  effective $\gamma$-ray 
detection in significantly broader intervals, 
namely each of them covers at least 2 decades in energy.
Thus the energy domains of these arrays largely overlap. 
Since all 4 versions of IACT arrays contain, from instrumental 
perspectives, the same  basic elements, and generally have  
common scientific motivations, an ideal arrangement would be the
combination of the \textit{sub-TeV, TeV, }and 
\textit{sub-PeV } (sub)arrays in a single 
array  with a quite homogeneous 
coverage (in the sense of performace) throughout the 
energy region from 
approximately 30 GeV to 300 TeV. The  
high $\gamma$-ray detection rates, coupled with 
good angular and energy resolutions over four 
energy decades would  make these combined arrays as 
multi-functional and multi-purpose  Ground Based Gamma-Ray 
Observatories (GBGROs) with a great capability for \textit{spectrometric,}
\textit{morphological} and \textit{temporal} studies  of a 
diverse range  of persistent and transient 
high energy phenomena in the Universe.  
There  is little doubt 
that the construction of such  
observatories,  desirably (at least) one 
in the Southern and another in the Northern Hemispheres, 
will lead to  many  unique  results and exciting discoveries.
The sites should have good optical  conditions and 
optimal,  close to 2km  asl, 
altitudes\footnote{Actually this should be considered as 
a \textit{compromised} rather than an optimal 
altitude, taking into account 
that while 2km elevation  is  optimal for TeV measurements, 
for sub-TeV and multi-TeV (sub-PeV) energy bands 
higher (more than 3km)  and lower (close to the sea level) 
elevations are more favorable.}. 
The \textit{HESS} site in Namibia,  as well as  the site of the 
Pierre Auger Cosmic Ray Observatory in Argentina in the 
Southern Hemisphere, and 
several sites in Northern Hemisphere (e.g. in Arizona (USA),  
Canary Islands La Palma or Tenerife, \textit{etc.}) 
match well these requirements.

\vspace{2mm}

The  \textit{sub-TeV} (sub)arrays  of future GBGROs 
with adequate  performance require  deeper design 
and technological studies,  albeit   some  30m diameter 
class prototype telescopes  are expected already 
in the foreseeable future, in particular  as integrates  of  
the plans of enlargement of the 
\textit{HESS}\cite{Punch-ch2005}
and \textit{MAGIC}\cite{Teshima-ch2005}  arrays. The operation of these single
dishes will  provide, together with GLAST,  the first reasonably 
deep probes of the sky  in the yet unknown  sub-100 GeV energy region. 

\vspace{2mm}

The main issue of realization of the  "10 $km^2$"  \textit{sub-PeV} 
(sub)arrays,  seems to be  related to the production of relatively cheap imagers
with large FoV  exceeding $5^\circ$. 
On shorter timescales, it would be important to build  
an independent sub-PeV  array in the Southern Hemisphere,
e.g. in  Australia which has suitable sites at low 
(close to the sea level)  altitudes which are beneficial in terms
of collection areas for $\geq 10$ TeV observations. Such an array  
consisting of  large number of  small aperture 
telescopes will be a powerful instrument in its own right
(designed  for  the discovery of "Cosmic PeVatrons"), 
\textit{the  fast scientific return  of which  seems to be quite secure, 
given the extension of energy spectra of many 
\textit{HESS} sources beyond several TeV.
}
\vspace{2mm}

While  all energy bands of future GBGROs are equally important and 
complementary,  the highest priority and preference 
(in terms of  the order of accomplishment) should be 
given (in my view) to  the $10^{-14} \ \rm erg/cm^2 s$  sensitivity 
\textit{TeV (sub)arrays}. One may predict with confidence that the 
construction of such a powerful
detector, which can be treated as a scaled-up version of  the current 
\textit{HESS} array, could be completed without unexpected complications
on relatively short  timescales. 
Such an array  does not require technological innovations, 
and relies on the approach, the reliability and feasibility of which  have  been clearly  demonstrated, although on smaller scales, 
by the \textit{CANGAROO, HESS, MAGIC}, and  {VERITAS} collaboration.  
In order to reduce the energy threshold 
to  $\approx  30$ GeV and  thus to  improve 
the flux sensitivity around 100 GeV, 
one may  consider somewhat  larger, e.g. 15m diameter
class telescopes  (but without compromising the field of view,
which should be as large as $5^\circ$), 
installed at  somewhat higher, 3 to 4 km  elevations 
\cite{Aharonian_Hamburg}. There are several suitable sites 
for such observatories in  both  Hemispheres, 
in particular in  northern Argentina and Chile.  
Two very attractive aspects of this 
option of  a GBGRO should be emphasized:  
(1) \textit{a very broad, from 30 GeV to 30 TeV
energy region  can be covered by
a single  array of identical telescopes with relatively 
modest optical reflectors, equipped with conventional 
PMT-based imagers, and installed at high but still 
comfortable altitudes}, and 
(2)  \textit{the design and construction 
of such a GBGRO can begin  right now}.

\vspace{2mm}

Although one may suggest to 
extend the energy domain  of  these GBGROs down
to $\leq 10$ GeV,  e.g. by adding  30m diameter class 
reflectors and/or using  high ($\geq 50 \%$) quantum efficiency 
imagers, I believe that the projects of future 
multi-GeV arrays should proceed through independent studies,
given the technological challenges (operation of 
large telescopes in robotic regime at high  altitudes,  
construction of high quantum efficiency focal plane imagers, \textit{etc.}). 
Because of  very  specific  astrophysical and cosmological 
goals, in particular the importance  of the  studies  of  highly variable $\gamma$-ray  phenomena in the 
remote  Universe (e.g. quasars, GRBs, \textit{etc.} at $z \geq 5$),
the reduction of the energy threshold to the lowest possible level 
is the  key issue for these \textit{Gamma-Ray Timing Explorers}.
That is why I believe that this activity should proceed through 
the  concept of 5@5.  The  successful realization  of 
a high-altitude IACT array 
\textit{during the  lifetime of GLAST} 
would be, of course, a great  achievement  of observational   
gamma-ray  astronomy\cite{Snowmass}.  
GLAST and 5@5 
are highly complementary instruments. While GLAST with its almost $2 \pi$ 
FoV can provide very effective monitoring of a very large  number of
sources, 5@5 has an obvious advantage for the
study of highly variable or transient $\gamma$-ray emitters. 
On the other hand, 5@5 is a detector with a small FoV, therefore 
it requires a special strategy of observations. Because of the 
overlap of the energy bands covered by these two instruments,
GLAST may serve  as  a perfect "guide" for 5@5. All sources, that will be detected by GLAST, can be potential target for observations 
with 5@5. 
The second, more strategic motivation for the activity 
towards  high altitude IACT arrays with energy threshold 
as low as several GeV is related to the lack of  any realistic 
satellite-based alternative  for GeV detectors in the the post-GLAST era.  In this regard \textit{the scientific reward of the 
implementation of ground-based approach in GeV gamma-ray astronomy
will be  enormous.}

\vspace{2mm}

Finally, a few comments concerning the large-field-of-view 
ground-based gamma-ray detectors. 
Three possible approaches have been proposed in this direction -
(i) very  large FoV  imaging air Cherenkov telescope
technique based on refractive optics \cite{Kifune2001},
(ii) arrays of $15^\circ$ FoV  IACTs\cite{Vassiliev-ch2005},  
and  (iii) dense  air shower particle arrays or large water Cherenkov
detectors installed at very high, $\geq 4$ km 
altitudes\cite{Sinnis-ch2005}.
While  the first  two techniques  require several technological  innovations,
the 3rd  approach  does not face serious technological  challenges. 
The feasibility of both the high altitude air shower array
and water Cherenkov  techniques  have been convincingly
demonstrated by the Tibet and Milagro collaborations
(see e.g. \cite{Snowmass}).  

\vspace{2mm}

The imaging atmospheric Cherenkov telescopes
are designed for  observations  of   $\gamma$-rays  from
objects with well determined positions. However, the 
high sensitivity of stereoscopic arrays coupled with relatively large
field-of-view homogeneous imagers  
may allow  quite effective  sky  surveys
as well,  as  has been convincingly demonstrated 
by the \textit{HESS} collaboration.
The next  generation GBGROs, in particular the  
$10^{-14} \ \rm erg/cm^2 s$ sensitivity  \textit{TeV}  (sub) arrays
will  provide much deeper  \textit{all-sky} (not limited by the 
galactic plane)  surveys. 
In particular,  all point $\gamma$-ray 
sources with fluxes at the  0.01 to 0.1Crab flux level
(depending on the energy band) 
can  be  be revealed  within several  steradians  of the sky during 
 one-year survey.

\vspace{2mm}

Nevertheless, the development of a  ground-based
technique allowing {\em simultaneous} coverage of
a significant  (1 steradian  or so) fraction of the sky is 
a high priority issue.   Actually the "standard" motivation for 
very large ground based gamma-ray  detectors 
-- \textit{all sky surveys} --   is  a reasonable  
but perhaps  not the strongest argument in favor of 
such  instruments.  In fact,  similar or even  deeper
\textit{all sky surveys} can be conducted with the 
next generation  IACT arrays. 
The strongest  motivation  of  the "1~steradian~FoV" detectors 
is, in fact,   their  unique potential allowing  
effective monitoring of $\gamma$-ray  activity of 
a large number of highly variable sources  like 
blazars  and microquasars , as well as 
the possibility for independent detection and study of solitary GeV-TeV 
$\gamma$-ray  events,  both \textit{related } (as  GeV-TeV counterparts) 
and \textit{not related} to classical (keV-MeV) GRBs. 
The particle acceleration and the secondary $\gamma$-ray  production processes in compact objects  proceed on extremely
short time-scales, often with main energy release 
in the  TeV band. Therefore the $\gamma$-ray emission 
carries unique information about the dynamics of these 
compact objects.  \textit{The promise  of exciting discoveries 
of  yet unknown VHE transient phenomena in the Universe 
fully  justifies  the efforts  towards  the 
construction of  large field-of-view   
ground-based gamma-ray 
detectors}.  Clearly,  these instruments will be complementary to GLAST 
and the  future large volume ("km$^3$" class) high energy neutrino detectors. 

\section*{Acknowledgments}
I thank Gavin Rowell and Dieter Horns for useful discussions.

%
\label{AharonianEnd}
 
\end{document}